\documentclass[12pt]{article}
\usepackage{amssymb,amsmath,epsfig}

\begin{document}

\title{\bf Phase Space Analysis for Anisotropic Universe with Nonlinear Bulk Viscosity}
\author{M. Sharif \thanks {msharif.math@pu.edu.pk} and Saadia Mumtaz
\thanks{sadiamumtaz17@gmail.com}\\
Department of Mathematics, University of the Punjab,\\
Quaid-e-Azam Campus, Lahore-54590, Pakistan.}

\date{}
\maketitle

\begin{abstract}
In this paper, we discuss phase space analysis of locally
rotationally symmetric Bianchi type I universe model by taking a
noninteracting mixture of dust like and viscous radiation like fluid
whose viscous pressure satisfies a nonlinear version of the
Israel-Stewart transport equation. An autonomous system of equations
is established by defining normalized dimensionless variables. In
order to investigate stability of the system, we evaluate
corresponding critical points for different values of the
parameters. We also compute power-law scale factor whose behavior
indicates different phases of the universe model. It is found that
our analysis does not provide a complete immune from fine-tuning
because the exponentially expanding solution occurs only for a
particular range of parameters. We conclude that stable solutions
exist in the presence of nonlinear model for bulk viscosity with
different choices of the constant parameter $m$ for anisotropic
universe.
\end{abstract}

\section{Introduction}

It is evident through many astronomical observations that our
universe is undergoing an accelerated expansion at its present
stage. This primal fact is supported by the observational probes of
various astronomical advances (type Ia supernova, large scale
structure and cosmic microwave background radiation (CMBR)) that
puts forward an open question on the current understanding of
fundamental physics \cite{1}. These observations suggest two cosmic
phases of accelerated expansion, i.e., the cosmic state before
radiation (the primordial inflationary era) and ultimately the
present cosmos phase after the matter dominated era. In the last
couple of decades, it is speculated that some mysterious source of
energy with unusual anti-gravitational force is responsible for the
current cosmic expansion dubbed as dark energy (DE).

The existence of this energy with large negative pressure does not
cluster at large scales. The study of the dominant constituents of
matter distribution in the universe has remained one of the most
debatable issues. Recent observations show that the visible part of
our universe is made up of baryonic matter contributing only $5\%$
of the total budget while the remaining ingredients yield the total
energy density composed of non-baryonic fluids ($68\%$ DE and $27\%$
dark matter) \cite{3}. The dark matter is an unusual material which
can be detected through its gravitational effects and neither emits
nor absorbs light \cite{4}.

In order to study the ambiguous nature of DE, several proposals have
been introduced in literature among them a small cosmological
constant ($\Lambda$) governed by a negative equation of state (EoS)
parameter ($\omega=-1$) is considered to be the simplest
characterization of DE. However, this identification has two
well-known problems, i.e., fine-tuning and cosmic coincidence. In
addition, there are several dynamical models which can be considered
as an alternative to $\Lambda$. These candidates involve scalar
field models like quintessence \cite{5}, phantom model \cite{6},
tachyon field \cite{7} and k-essence \cite{8} that also suggest
expanding behavior of the universe. Another approach involves the
generalization of simple barotropic EoS to more exotic forms such as
Chaplygin gas \cite{9} and its modification \cite{10}.

It has been suggested that a fluid with bulk viscosity may cause an
accelerated expansion of the universe models without cosmological
constant or scalar field \cite{10a}. The bulk viscosity refers to
the measure of pressure required to restore an equilibrium state
when cosmic expansion of any fluid occurs in an expanding universe
scenario. In case of thermodynamics, bulk viscosity occurs due to
its deviation from local thermodynamical equilibrium in any physical
system \cite{10b}. Our main concern is to explore another approach
which tends to minimize the exotic forms of matter by introducing
dissipation through viscous effects of fluids. Bulk viscous pressure
provides the dissipative contribution which has a significant
relevance in homogeneous universe scenarios.

A phase space is a space describing all possible states (position
and momentum) corresponding to each point of the system. The study
of possible stable late-time attractors has attained remarkable
significance for different universe models. A phase space analysis
manifests dynamical behavior of a cosmological model through a
global view by reducing complexity of the equations (converting the
system of equations to an autonomous system). This analysis is
helpful to comprehend different patterns of evolution. A linearly
stable fixed point will behave as an attractor for the neighboring
points which ultimately leads to converging trajectories. This
analysis only deals with the stability of any system by checking
whether the system remains stable for a long time or the initial
data has any impact \cite{11}. The study of stability of different
universe models via phase space portraits helps to explore their
qualitative features.

Copeland et al. \cite{12} discussed a phase plane analysis of
standard inflationary models and analyzed that these models cannot
solve density problem. Guo et al. \cite{13} studied phase space
analysis for FRW universe model filled with barotropic fluid and
phantom scalar field in which phantom dominated solution is found as
a stable late-time attractor. Yang and Gao \cite{14} explored phase
space analysis of k-essence cosmology and found that stability of
model as well as critical points play an important role for the
final state of the universe. Xiao and Zhu \cite{15} analyzed
stability of FRW universe model in loop quantum gravity by using
phase space analysis along with barotropic fluid and positive field
potential. Acquaviva and Beesham \cite{16} discussed phase space
analysis by taking FRW spacetime filled with noninteracting mixture
of fluids (dust and viscous radiation) and found that the nonlinear
viscous model describes the possibility of current accelerated
expansion of the universe. They extended this dynamical analysis by
taking three dimensionless variables in the context of non-viscous
dust and viscous radiation \cite{17}. We discussed the impact of
nonlinear electrodynamics on stability of accelerated expansion of
FRW universe model with nonlinear bulk viscosity \cite{17a}.

Bianchi universe models are considered to be appropriate for the
cosmological description of various states of the expanding
universe. These models are widely discussed in literature to study
expected primordial anisotropy and some large angle anomalies
detected by CMBR which yield violation of statistical isotropy of
cosmos \cite{18}. Coley and Dunn \cite{19} used phase plane
techniques to study dynamical behavior of Bianchi type V model
containing a viscous fluid. Sharif and Waheed \cite{20} explored
phase space analysis of locally rotationally symmetric (LRS) Bianchi
type I (BI) universe for chameleon scalar field in Brans-Dicke
gravity. Sharif and Jabbar \cite{21} studied stability of LRS BI
universe model through phase space analysis for phantom, non-phantom
and vacuum phases in generalized teleparallel gravity. Recently, we
have investigated the phase phase analysis of LRS BI universe model
coupled with linear bulk viscosity \cite{21a} and phantom as well as
tachyon models \cite{21b}.

The theme of this paper is to study the phase space analysis of LRS
BI universe with nonlinear viscous fluid. The plan of the paper is
as follows. In section \textbf{2}, we provide some basic equations
and a nonlinear model for bulk viscosity. In order to analyze
stability of the system, an autonomous system of equations is
established by introducing normalized dimensionless variables in
section \textbf{3}. Section \textbf{4} deals with the formulation of
power-law scale factor. Finally, we provide a brief overview of the
obtained results in the last section.

\section{General Equations}

Bianchi universe models are considered to be appropriate for the
cosmological description of various states of expanding universe. It
has been observed that some large angle anomalies in CMBR tend to
violate the statistical isotropy of present cosmic models \cite{18}.
In this context, homogeneous anisotropic universe models under plane
symmetric background play a significant role for the better
understanding of these anomalies. The LRS BI model is the simplest
generalization of FRW universe by adding effects of anisotropy so it
would be interesting to explore the stability of this model via
phase space analysis. The line element for LRS BI is given by
\cite{23}
\begin{equation}\label{1}
ds^2=-dt^2+a^2(t)dx^2+b^2(t)(dy^2+dz^2),
\end{equation}
where $a(t)$ and $b(t)$ represent cosmic expansion radii. The
corresponding mean Hubble parameter is defined as
\begin{equation}\label{1a}
H=\frac{1}{3}[H_{1}+H_{2}]=\frac{1}{3}\left[\frac{\dot{a}}{a}
+\frac{2\dot{b}}{b}\right],
\end{equation}
where $H_{1}=\frac{\dot{a}}{a}$ and $H_{2}=\frac{\dot{b}}{b}$ are
directional Hubble parameters. We can define the expansion scalar
through scale factors as
\begin{equation}\nonumber
\Theta=\left[\frac{\dot{a}}{a}+\frac{2\dot{b}}{b}\right].
\end{equation}

In case of LRS BI model, we obtain Raychaudhuri and constraint
equations from the field equations which lead to dynamical system of
equations. These equations are quite complicated due to the presence
of two scale factors (the number of equations is less than the
number of unknown parameters). In order to reduce the complexity of
the system, we require an additional constraint relating these
parameters so that we can obtain explicit solution of the system.
For a spatially homogeneous spacetime, the normal congruence to
homogeneous expansion leads to a constant ratio, i.e., the expansion
and shear scalars are proportional to each other \cite{aa}. For LRS
BI model, its integration leads to the condition $a=b^m$,
$m\neq0,1$, where $m$ is a constant parameter such that LRS BI model
reduces to homogeneous and isotropic (FRW) universe model for the
case $m=1$. A relationship between mean and directional Hubble
parameters, representing the average Hubble expansion in one
direction, is given as
\begin{equation}\label{1b}
H_{1}=mH_{2}=\left(\frac{3m}{m+2}\right)H.
\end{equation}
The physical reason for this assumption is justified by the
observations of the velocity redshift relation for extragalactic
sources which suggest that the Hubble expansion of the universe may
achieve isotropy when shear to expansion scalar ratio is constant
\cite{d}. Collins \cite{dd} discussed physical significance of this
condition for perfect fluid and barotropic EoS in a more general
case. This condition has been used by many authors in literature
\cite{d2}. An anisotropic model with the diagonal energy-momentum
tensor may yield isotropic universe in the limit $t\rightarrow
+\infty$ and positive energy density. Collins and Hawking \cite{c}
described the criterion for having the possibility of such a model
where they established that the anisotropy vanishes in the limit
$t\rightarrow\infty$.

In the framework of homogenous and anisotropic spacetimes, it has
generally been assumed that cosmic fluid yields isotropic pressure.
Various discussions have promoted the general interest not only in
the Bianchi type cosmological models but also in the possibility of
anisotropic nature of cosmic fluid \cite{23a1}. Bianchi universe
models can admit both isotropic as well as anisotropic pressure
depending upon the chosen matter distribution. In our case, we are
dealing with the simple case by considering isotropic fluid. The
matter distribution for the cosmic fluid is given by
\begin{equation}\nonumber
T_{\alpha\beta}=(\sigma+p)u_{\alpha}u_{\beta}+pg_{\alpha\beta},
\end{equation}
where $\sigma$, $p$ and $u_{\alpha}$ correspond to the energy
density, total pressure and four-velocity, respectively. We consider
that the universe model is filled with two fluids, i.e., a
noninteracting dust like fluid with energy density
$\sigma_{d}(p_{d}=0)$ and a viscous radiation like fluid having
energy density $\sigma_{v}$ as well as the effective pressure
$p=p_{v}(\sigma_{v})+\Phi$ \cite{16,17,b,23a}. Here $p_{v}$
corresponds to the normal or equilibrium pressure for which we
assume a barotropic EoS as a viscous fluid given by
\begin{equation}\label{6}
p_{v}=(w-1)\sigma_{v},
\end{equation}
where $1\leq w\leq2$ is the EoS parameter. Also, $\Phi$ is the
non-equilibrium part, i.e., bulk viscous pressure satisfying an
evolution equation. The main contribution of bulk viscosity to the
effective pressure includes its dissipative effect. It is mentioned
here that the model under consideration admits both types of
viscosity (bulk and shear) with bulk viscosity being the dominant
dissipative stress only in the radiative mixture of non-relativistic
baryons and radiation \cite{b1}. So we cannot rule out the shear
viscosity in the respective fluid but can assume the dominance of
bulk viscosity in our case. Bulk viscosity arises typically in
mixtures either of different species as in a radiative fluid or of
the same species but with different energies as in a
Maxwell-Boltzmann gas. Physically, we can think of bulk viscosity as
the internal friction that sets in due to different cooling rates in
the expanding mixture. The Raychaudhuri equation obtained from the
Einstein field equation is given by
\begin{equation}\label{2}
\dot{\Theta}=-\frac{1}{m+2}\Theta^2-\frac{m+2}{2}\left[\frac{1}{2m+1}
(\sigma_{d}+\sigma_{v})+p_{v}+\Phi\right],
\end{equation}
where dot means derivative with respect to time. The constraint
equation yields
\begin{equation}\label{3}
\sigma_{d}+\sigma_{v}-\frac{2m+1}{(m+2)^2}\Theta^2=0,
\end{equation}
which enables us to consider only the evolution of viscous energy
density without dust component. The conservation of energy-momentum
tensor leads to the following evolution equations for viscous and
dust components
\begin{eqnarray}\label{4}
\dot{\sigma}_{v}&=&-[\sigma_{v}+p_{v}+\Phi]\Theta, \\\label{5}
\dot{\sigma}_{d}&=&-\sigma_{d}\Theta.
\end{eqnarray}

Using Eqs.(\ref{2}) and (\ref{3}), Raychaudhuri and conservation
equations for viscous fluid become
\begin{eqnarray}\label{7}
\dot{\Theta}&=&-\frac{3}{2(m+2)}\Theta^2-\frac{m+2}{2}[(w-1)\sigma_{v}+\Phi],
\\\label{8} \dot{\sigma}_{v}&=&-[w\sigma_{v}+\Phi]\Theta.
\end{eqnarray}
The viscous pressure variable can be characterized by an evolution
equation given by \cite{b}
\begin{equation}\label{9}
\tau\dot{\Phi}=-\zeta\Theta-\Phi\left(1+\frac{\tau_{*}}
{\zeta}\Phi\right)^{-1}-\frac{1}{2}\tau\Phi\left[\Theta+\frac{\dot
{\tau}}{\tau}-\frac{\dot{\zeta}}{\zeta}-\frac{\dot{T}}{T}\right],
\end{equation}
where $\zeta$, $T$, $\tau$ and $\tau_{*}$ represent bulk viscosity,
local equilibrium temperature, linear relaxation time and
characteristic time in nonlinear background, respectively. This
equation is derived by using a nonlinear model describing a
relationship between thermodynamic flux ``$\Phi$'' and thermodynamic
force ``$\chi$" in the form
\begin{equation}\label{10}
\Phi=-\frac{\zeta\chi}{1+\tau_{*}\chi}.
\end{equation}
This is a nonlinear extension of Israel-Stewart equation which
reduces to its linear form as $\tau_{*}\rightarrow0$. The nonlinear
term in Eq.(\ref{9}) must be positive for thermodynamic consistency
and positivity of entropy production rate. It can be speculated that
the respective fluid can be a gas of unknown non-relativistic or
ultra-relativistic particles having thermodynamic parameters
$\zeta$, $\tau$, $\tau_{*}$ and $T$. We need to specify these
thermodynamic parameters as follows. The parameter for
characteristic time which gives the qualitative nature of nonlinear
effects is defined as \cite{b2}
\begin{equation}
\tau_{*}=k^2\tau,
\end{equation}
where $k$ is a dimensionless constant. This mathematical assumption
allows us to analyze some qualitative features of nonlinear bulk
viscosity. The linear relaxation time can be related to the bulk
viscosity by the following relation
\begin{equation}
\tau=\frac{\zeta}{wv^2\sigma_{v}},
\end{equation}
where $v$ corresponds to the dissipative effect of the speed of
sound $V$ such that $V^2=c^2_{s}+v^2$, where $c^2_{s}$ is its
adiabatic contribution. By causality, $V\leq1$ and $c^2_{s}=w-1$
which yields
\begin{equation}\label{11}
v^2\leq2-w, \quad 1\leq w\leq2.
\end{equation}
We can define the bulk viscosity in terms of expansion scalar as
\begin{equation}
\zeta=\zeta_{0}\Theta,
\end{equation}
with $\zeta_{0}>0$ as a constant. We also express temperature of the
system as barotropic temperature $T=T(\sigma)$ given by
\begin{equation}
T=T_{0}\sigma^{(w-1)/w}.
\end{equation}
The explicit form of evolution equation in the context of above
relations leads to
\begin{equation}\label{12}
\dot{\Phi}=-wv^2\sigma_{v}\Theta-\frac{wv^2\Phi\sigma_{v}}{\zeta_{0}\Theta}
\left(1+\frac{k^2\Phi}{wv^2\sigma_{v}}\right)^{-1}-\frac{1}{2}
\Phi\left[\Theta-\left(\frac{2w-1}{w}\right)\frac{\dot{\sigma_{v}}}
{\sigma_{v}}\right].
\end{equation}

\section{Phase Space Analysis}

This section deals with phase space analysis of LRS BI universe
model for dust like and viscous radiation like fluids. Due to many
arbitrary parameters, it seems difficult to find analytical solution
of the evolution equation. For this purpose, we define normalized
dimensionless variables
$\Omega=\frac{3(m+2)\sigma_{v}}{(2m+1)\Theta^2}$ and
$\tilde{\Phi}=\frac{3(m+2)\Phi}{(2m+1)\Theta^2}$ which can reduce
this dynamical system to an autonomous one, where $m\neq-1/2$ to
avoid singularity. We also define a new variable
$\frac{dt}{d\tau}=\frac{3}{\Theta}$ for time through which the
corresponding derivative will be represented by prime. Here each
term is associated with some physical explicit origin since the
chosen dimensionless variables $\Omega$ and $\tilde{\Phi}$ occur due
to physical impact of viscous energy density and pressure,
respectively. The system of Eqs.(\ref{7}) and (\ref{8}) in terms of
these normalized variables become
\begin{eqnarray}\label{13}
\frac{\Theta'}{\Theta}&=&-\frac{3}{2}\left[\frac{3}{m+2}
+\frac{2m+1}{3}[(w-1)\Omega+\tilde{\Phi}]\right],\\\label{14}
\frac{3\sigma_{v}'}{\Theta^2}&=&-\frac{3(2m+1)}{m+2}[w\Omega+\tilde{\Phi}].
\end{eqnarray}
The dimensionless variable for energy density, i.e., $\Omega$ yields
\begin{equation}\label{15}
\Omega'=\frac{m+2}{2m+1}\left[\frac{3\sigma_{v}'}{\Theta^2}
-2\Omega\frac{\Theta'}{\Theta}\right].
\end{equation}
Using Eqs.(\ref{13}) and (\ref{14}), this equation turns out to be
\begin{equation}\label{16}
\Omega'=[(m+2)\Omega-3][\Omega(w-1)+\tilde{\Phi}].
\end{equation}
The first derivative of $\tilde{\Phi}$ with respect to $\tau$
through Eq.(\ref{13}) leads to an evolution equation of the form
\begin{eqnarray}\nonumber
\tilde{\Phi}'&=&-wv^2\Omega\left[1+\frac{2m+1}{m+2}\frac{\tilde{\Phi}}
{3\zeta_{0}}\left(1+\frac{k^2\tilde{\Phi}}{wv^2\Omega}\right)^{-1}\right]
-\frac{\tilde{\Phi}}{2}\left(1+\frac{2w-1}{w(m+2)}\right)\\\nonumber&+&
(2m+1)(w-1)\left[1-\frac{2w-1}{2w}\frac{\tilde{\Phi}}{9}\right]
+(2m+1)\tilde{\Phi}^2\left[1-\frac{2w-1}{2w}\right].\\\label{17}
\end{eqnarray}
It is mentioned here that Eqs.(\ref{16}) and (\ref{17}) have a
substantial role to describe the dynamical system under
consideration for phase space analysis. In order to find the
critical points $\{\Omega_{c},\tilde{\Phi}_{c}\}$, we need to solve
the respective dynamical system by imposing the condition
$\Omega'=\tilde{\Phi}'=0$. The stability of LRS BI universe model
will be examined according to the nature of critical points.

In order to find a region corresponding to the accelerated
expansion, we follow \cite{b}. In this context, we define the
entropy four-current in the form
\begin{equation}\label{1aa}
S^{\alpha}=S_{eff}n^{\alpha},
\end{equation}
where $S_{eff}$ represents the effective specific entropy. Also, the
particle number four-current is given by
\begin{equation}\label{2aa}
n^{\alpha}=nu^{\alpha},
\end{equation}
whose conservation equation yields
\begin{equation}\label{3aa}
\dot{n}=-\Theta n.
\end{equation}
In Israel-Stewart theory, we have
\begin{equation}\label{6aa}
S_{eff}=S-\left(\frac{\tau}{2nT\zeta}\right)\Phi^2.
\end{equation}
The local equilibrium variables $S$ and $T$ satisfy the Gibbs
equation as follows
\begin{equation}\label{7aa}
TdS=(\sigma_{v}+p_{v})d\left(\frac{1}{n}\right)+\frac{1}{n}d\sigma_{v},
\end{equation}
which, through Eqs.(\ref{4}) and (\ref{3aa}), implies that
$\dot{S}=-\frac{\Theta\Phi}{nT}$. Equations (\ref{1aa}) and
(\ref{6aa}), through (\ref{4}), (\ref{3aa}) and (\ref{7aa}), give
\begin{equation}\label{9aa}
S^{\alpha}_{~;\alpha}=-\frac{\Phi\chi}{T}.
\end{equation}
Using Eqs.(\ref{10}) and (\ref{9aa}), we find
\begin{equation}\label{10aa}
S^{\alpha}_{~;\alpha}=n\dot{S}_{eff}=\frac{\Phi^2}{\zeta
T}\left[1+\frac{\tau_*\Phi}{\zeta}\right]^{-1}.
\end{equation}

The second law of thermodynamics yields positivity of entropy rate
given by
\begin{equation}\nonumber
S^{\alpha}_{~;\alpha}\geq0,
\end{equation}
such that the second law holds identically by virtue of the upper
bound on the bulk stress as
\begin{equation}\label{11aa}
\tilde{\Phi}\geq-\frac{\zeta}{\tau_*}.
\end{equation}
If $\tilde{\Phi}=-\frac{\zeta}{\tau_*}$, the entropy production rate
becomes undefined due to the inverse term. Thus we restrict the
phase space region to a condition necessary for the positivity of
entropy production rate which demands \cite{16}
\begin{equation}\label{18}
\tilde{\Phi}>-\frac{wv^2\Omega}{k^2}.
\end{equation}
This condition tends the possible negative values of $\tilde{\Phi}$
towards zero for $k^2\gg v^2$. Contrarily, the bulk pressure will be
less restrictive if $k^2\ll v^2$. In the limit $v\rightarrow0$,
finite values of $k$ allow only positive values of bulk pressure. It
would be more convenient to consider $k^2\leq v^2$ along with
$v^2\leq2-w$ and $\tau_{*}=k^2\tau$ which implies that the
characteristic time for nonlinear effects $\tau_{*}$ does not exceed
the characteristic time for linear background $\tau$. The critical
points can be characterized by some important quantities which
include deceleration parameter $q=-1-\frac{\Theta'}{\Theta}$ and
effective EoS parameter $w_{eff}=-\frac{2\Theta'}{3\Theta}$ leading
to
\begin{eqnarray}\label{19}
q&=&\frac{1}{2}\left[\frac{5-2m}{m+2}+(2m+1)\{(w-1)\Omega+\tilde{\Phi}\}\right],
\\\label{20}
w_{eff}&=&\frac{3}{m+2}+\frac{2m+1}{3}[(w-1)\Omega+\tilde{\Phi}].
\end{eqnarray}

In order to explore a region of phase space undergoing accelerated
expansion, we impose $q<0$ in Eq.(\ref{19}) which yields
\begin{equation}\label{21aa}
\tilde{\Phi}<\frac{2m-5}{(2m+1)(m+2)}-(w-1)\Omega.
\end{equation}
The possibility of accelerated expansion in the physical phase space
is determined by comparing Eqs.(\ref{18}) and (\ref{19}) through
$q<0$ given by
\begin{equation}\label{21a}
\frac{v^2}{k^2}>\frac{(2m+1)(m+2)(w-1)\Omega-2m+5}{(2m+1)(m+2)w\Omega}.
\end{equation}
Inserting $\Omega'=0$ in Eq.(\ref{16}), we identify the following
conditions
\begin{eqnarray}\label{21}
\Omega_{c}&=&\frac{3}{m+2},\\\label{22}
(w-1)\Omega_{c}+\tilde{\Phi}_{c}&=&0.
\end{eqnarray}
To locate the critical points, we need to insert these conditions in
$\tilde{\Phi}'$. This analysis is carried out by characterizing the
viscous fluid through the choice of its EoS parameter $w$ (dust or
radiation). We consider $0<k^2=v^2\leq2-w$ for which the case of
stiff matter ($w=2$) is excluded from the analysis because it will
give $v^2=0$. Here we provide a brief overview to this analysis as
follows.
\begin{itemize}
\item Convert the dynamical system of equations to the autonomous
system by using dimensionless variables.
\item Evaluate the nature of critical points
$\{\Omega_{c},\tilde{\Phi}_{c}\}$ of the above autonomous system to
discuss stability of model.
\item Calculate the eigenvalues of the Jacobi matrix which can
characterize these critical points.
\end{itemize}

\subsection{Dust Like EoS ($w=1$)}

We first consider the case of dust like fluid by taking $w=1$ for
phase space analysis. By taking the first condition
$\Omega_{c}=\frac{3}{m+2}$ and $\tilde{\Phi}'=0$ in Eq.(\ref{17}),
we have
\begin{eqnarray}\nonumber
&&\frac{v^2(2m+1)}{9\zeta_{0}}\tilde{\Phi}^3-(2m+1)\left(\frac{1}{2}
+\frac{v^2}{3\zeta_{0}(m+2)}\right)\tilde{\Phi}^2\\\label{23}&+&
\left(\frac{v^2}{\zeta_{0}}\frac{2m+1}{(m+2)^2}\right)\tilde{\Phi}
+\frac{3v^2}{m+2}=0.
\end{eqnarray}
This cubic equation yields three roots (two real and one imaginary)
for the considered parameters. Here we are concerned with real roots
$\tilde{\Phi}^+$ (positive) and $\tilde{\Phi}^-$ (negative) whose
corresponding critical points are $P^+_{d}$ and $P^-_{d}$,
respectively. It is mentioned here that the most negative root
always lies in the region of negative entropy production rate. The
general form of the dynamical system is given by
\begin{equation}\label{24}
\Omega'=f(\Omega, \tilde{\Phi}),\quad \tilde{\Phi}'=g(\Omega,
\tilde{\Phi}).
\end{equation}
The eigenvalues of the system can be determined by the Jacobian
matrix
\begin{equation}
A=\left(
\begin{array}{cc}
\frac{\partial f}{\partial\Omega}&\frac{\partial f}{\partial\tilde{\Phi}}\\
\frac{\partial g}{\partial\Omega}&\frac{\partial g}{\partial\tilde{\Phi}}\\
\end{array}
\right)_{|P_i^{\pm}}.
\end{equation}
The corresponding eigenvalues are given by
\begin{eqnarray}\label{26}
\lambda_{1}&=&\frac{\partial
f}{\partial\Omega}|_{P_{d}^{\pm}}=(m+2)\tilde{\Phi}^{\pm},
\\\label{27} \lambda_{2}&=&\frac{\partial
g}{\partial\Omega}|_{P_{d}^{\pm}}=(2m+1)\left[\tilde{\Phi}^{\pm}
-\frac{9v^2}{\zeta_{0}(m+2)^2[3+(m+2)\tilde{\Phi}^{\pm}]^2}\right].
\end{eqnarray}

The fixed point is called a source (respectively, a sink) if both
eigenvalues consist of positive (respectively negative) real parts.
The real parts of the eigenvalues having opposite signs correspond
to a saddle point of the system. The sign of both eigenvalues will
be positive for the point $P^+_{d}$ with $\tilde{\Phi}>0$ showing a
source (unstable). Also, the point $P_{d}^-$ with negative
eigenvalues corresponds to a stable sink. We consider the condition
(\ref{22}) which gives $\tilde{\Phi}=0$ in the case of dust like
fluid. This represents a line of the points where the flow is at
rest in the phase space region. However, the stability of point
$P_{0}^{0}=\{0,0\}$ is analyzed at the line where the entropy
production rate diverges and the system is not well defined. We are
interested to investigate the impact of $m$ on stability of the
critical points in the presence of nonlinear bulk viscosity. We plot
the dynamical behavior of critical points for different values of
$m$ corresponding to the dust case as shown in Figures \textbf{1}
and \textbf{2}.

In these numerical plots, the green trajectory represents a flow
from the point $P^{+}_{d}$ towards $P^{-}_{d}$ while the red
trajectory is a constraint which makes the phase space bounded in
$\Omega$ direction corresponding to different values of $m$ beyond
which the trajectories are not considered physically relevant. The
white region in the bottom shows the universe models with a negative
entropy production rate whereas this rate diverges on its boundary.
It is mentioned here that trajectories in the neighborhood of this
boundary are not attracted towards it showing its repulsive
behavior. This feature has remarkable significance to keep the
models away from divergence of the entropy production rate. The
green region corresponds to $q<0$ showing accelerated expansion of
the universe.
\begin{figure}\center
\epsfig{file=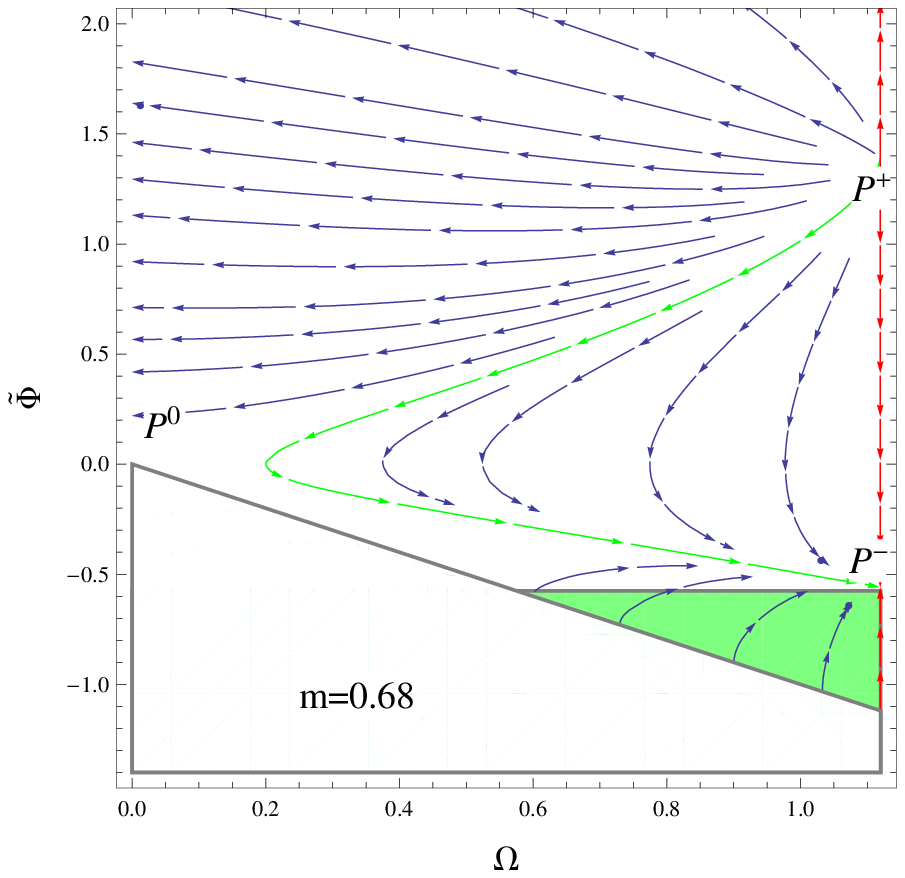,width=0.35\linewidth}\epsfig{file=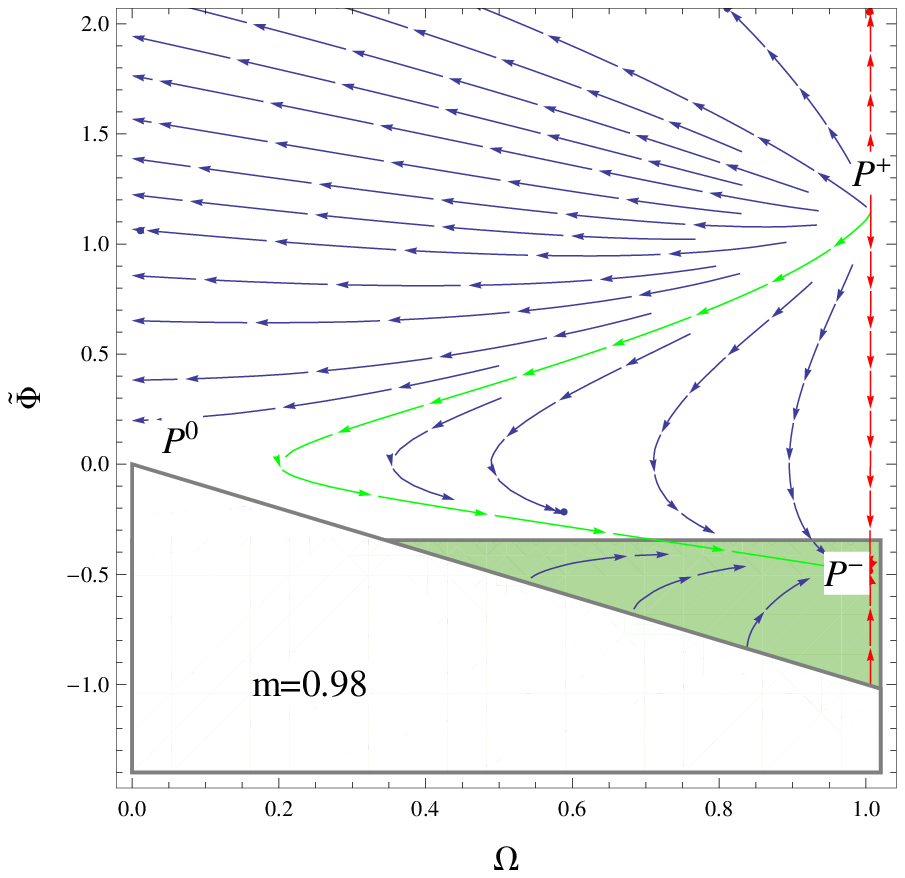,
width=0.35\linewidth}\epsfig{file=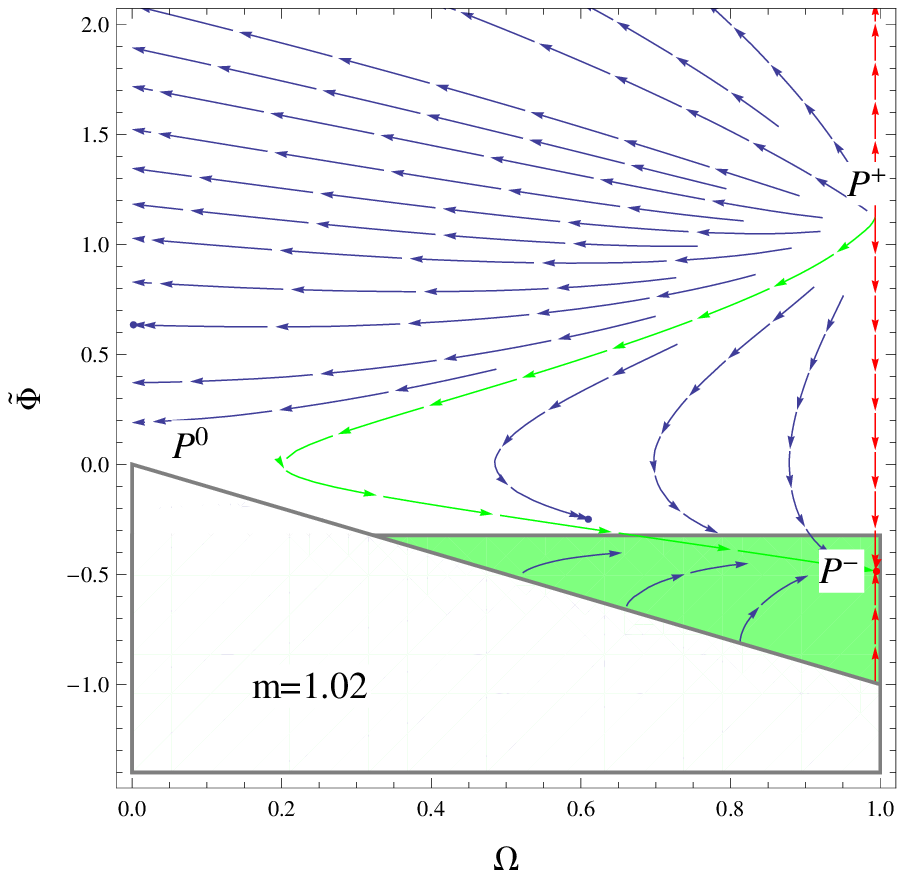,width=0.35\linewidth}
\caption{Plot for the phase plane evolution of LRS BI model with
$w=1$, $v^2=k^2=1$, $\zeta_{0}=1$ and different values of $m$. White
and green regions represent negativity of entropy production rate
and accelerated expansion, respectively.}
\end{figure}

We find that the green region (accelerated expansion) increases by
increasing $m$. For $m=0.68$ and $v^2=k^2=1$, the point $P_{d}^-$ is
a global attractor which lies in the physical phase space outside
the green region showing decelerated expansion of the universe model
dominated by matter. For $m\rightarrow1$, i.e., $m=0.98,~1.02$, it
is found that the global attractor $P_{d}^-$ lies in the green
region showing an expanding model (due to viscosity effects)
dominated by matter. The respective analysis is shown in Figure
\textbf{1}. For $v^2=k^2=0.04$, the graphical results show
decelerated expansion coming from both viscous radiation and
non-viscous dust for all choices of $m$ (Figure \textbf{2}). The
summary of the results for the stability of LRS BI model filled by
dust fluid is given in Table \textbf{1}.
\begin{table}[bht]
\textbf{Table 1:} \textbf{Stability Analysis of Critical Points for
Dust Case} \vspace{0.5cm} \centering
\begin{small}
\begin{tabular}{|c|c|c|c|}
\hline\textbf{Critical Point}&\textbf{$P^{0}_{d}$}&
$P^{-}_{d}$&$P^{+}_{d}$\\
\hline{\textbf{Behavior}}&{Saddle}&{Sink}&{Source}\\
\hline{\textbf{Stability}}&{Unstable}&{Stable}&{Unstable}\\
\hline
\end{tabular}
\end{small}
\end{table}
\begin{figure}\center
\epsfig{file=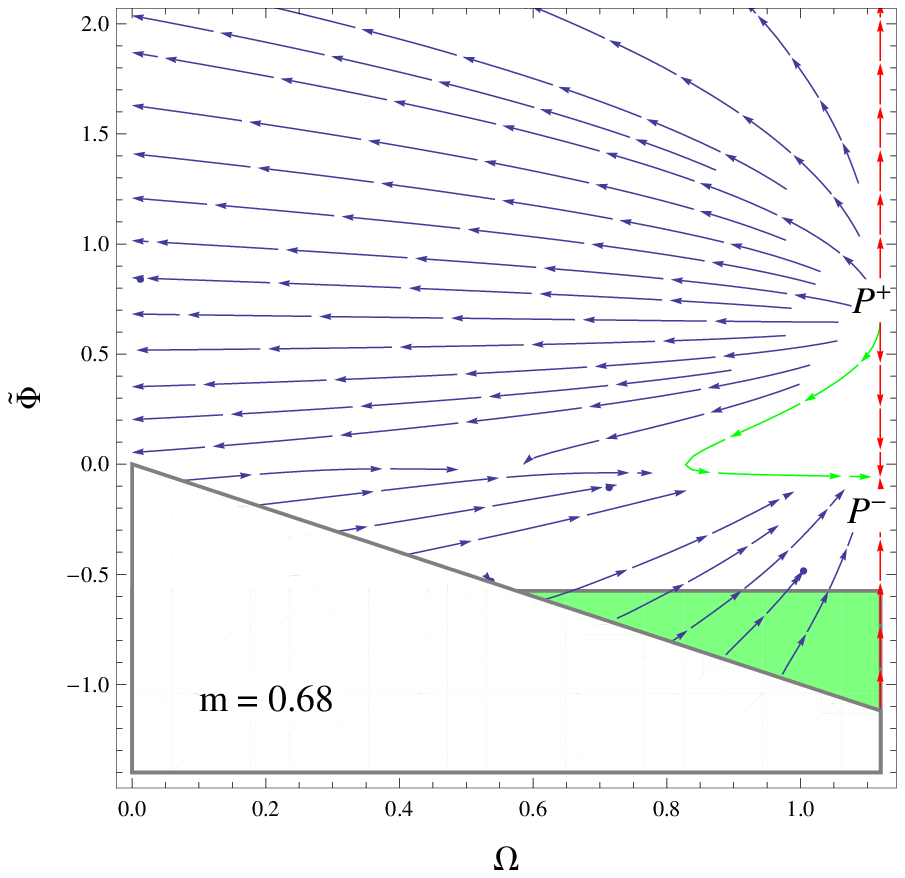,width=0.35\linewidth}\epsfig{file=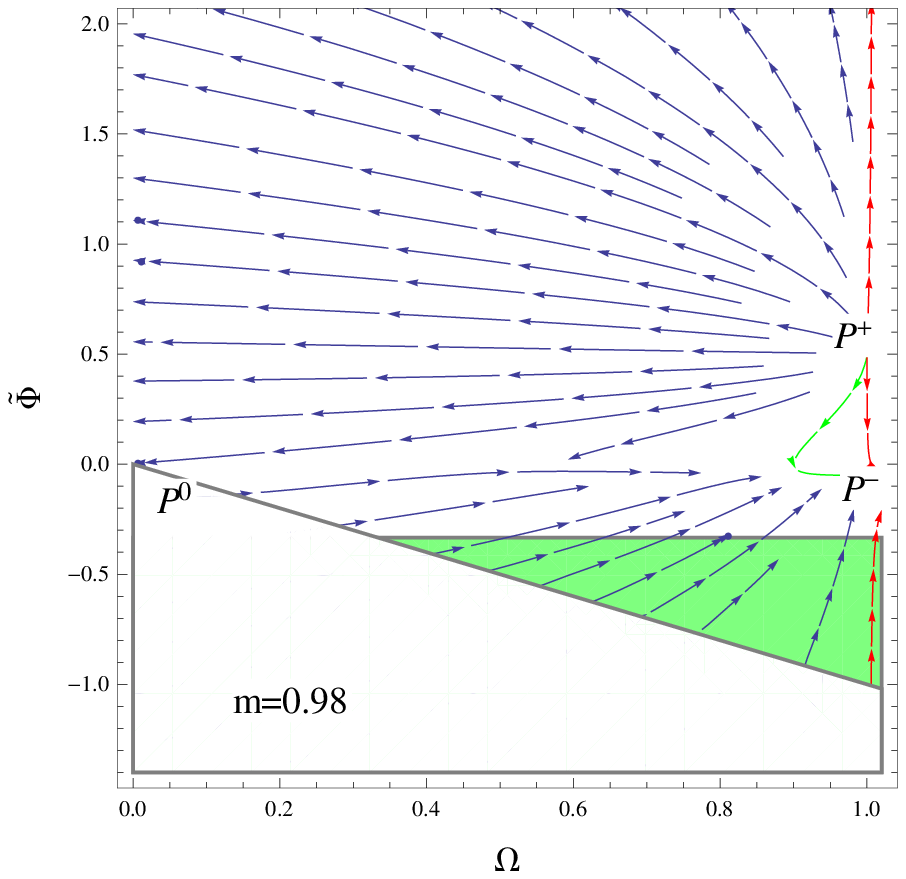,
width=0.35\linewidth}\epsfig{file=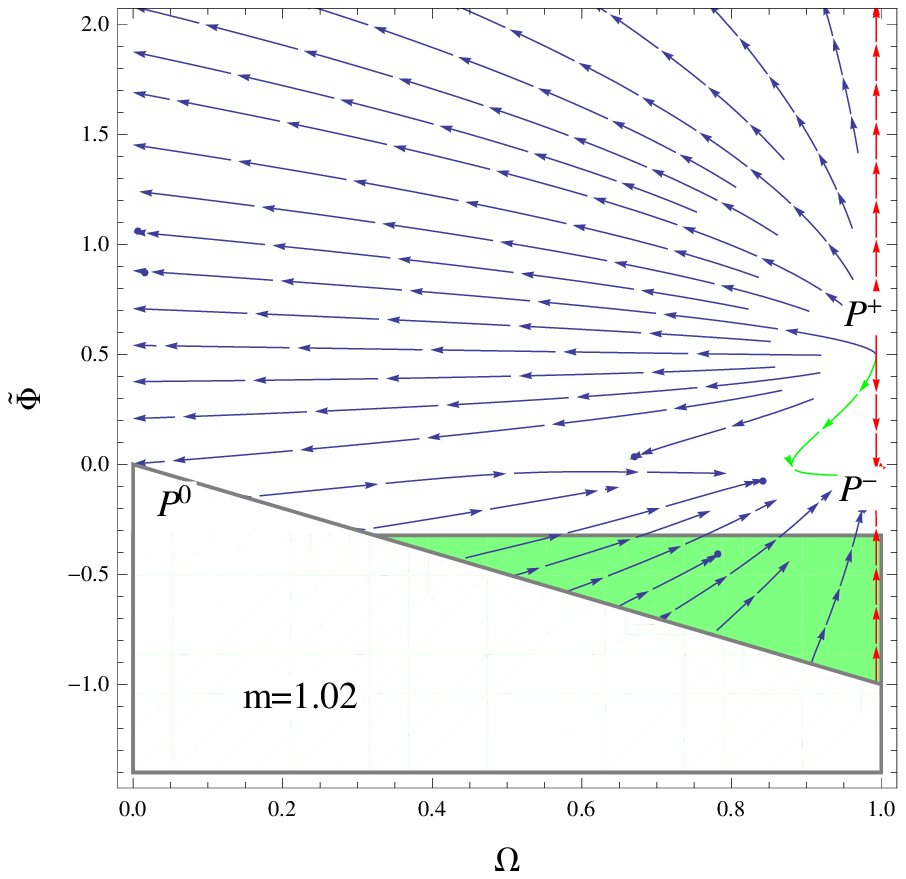,width=0.35\linewidth}
\caption{Plot for the phase plane evolution with $w=1$,
$v^2=k^2=0.04$, $\zeta_{0}=1$ and and different values of $m$.}
\end{figure}

\subsection{Radiation Like EoS $(w=\frac{4}{3})$}

Here we impose the condition (\ref{21}) and $\tilde{\Phi}'=0$ in
Eq.(\ref{17}) which gives a cubic equation of the form
\begin{eqnarray}\nonumber
&&\frac{v^2}{3\zeta_{0}}\tilde{\Phi}^3-\left(m+2
+\frac{4v^2}{\zeta_{0}}\right)\tilde{\Phi}^2+(m+2)(2m+1)
\left[\frac{2(2m+1)v^2}{\zeta_{0}}\right.\\\label{27}&&
+\left.\frac{1}{576}\{368m^2+1505m+1502\}\right]\tilde{\Phi}
+\frac{4v^2}{2m+1}-1=0.
\end{eqnarray}
This equation provides three roots among which we retain only those
roots that lie in the physical phase space. We find two critical
points $P_{r}^+=\{\frac{3}{m+2}, \tilde{\Phi}_{c}^+\}$ and
$P_{r}^-=\{\frac{3}{m+2}, \tilde{\Phi}_{c}^-\}$ corresponding to
positive $(\tilde{\Phi}_{c}^+)$ and negative $(\tilde{\Phi}_{c}^-)$
roots, respectively. Using $\tilde{\Phi}'=0$ and the second
condition (\ref{22}) with $\tilde{\Phi}_{c}=-\frac{\Omega_{c}}{3}$,
we obtain two critical points $P^0_{r}=\{0,0\}$ and
\begin{eqnarray}\nonumber
P^{*}_{r}&=&\left\{\frac{48v^2(m+2)\zeta_{0}}{(2m+1)
[128-9(m+2)\zeta_{0}]},-\frac{16v^2(m+2)\zeta_{0}}{(2m+1)
[128-9(m+2)\zeta_{0}]}\right\},\\\label{27a}
\end{eqnarray}
subject to the condition $\Omega_{c}\leq\frac{3}{m+2}$ for their
influence in the physical phase space region such that
\begin{eqnarray}\label{27aa}
0<v\leq \bar{v}, \quad\zeta_{0}>0, \quad
\zeta_{0}>\frac{1}{3(m+2)(m+1)^2}\bar{\zeta_{0}},
\end{eqnarray}
where
$\bar{v}=\frac{1}{2(m+2)}\sqrt{\frac{1}{6\zeta_{0}}[3(m+2)(m+1)^2
\zeta_{0}+\bar{\zeta_{0}}]}$ and $\bar{\zeta_{0}}=\frac{2m+1}{32}$.

The eigenvalues for stability matrix corresponding to the points
$P^{\pm}_{r}$ are given by
\begin{eqnarray}\label{28a}
\lambda_{1}&=&\frac{2}{3}(m+2)\Omega+(m+2)\tilde{\Phi}-1,
\\\nonumber\lambda_{2}&=&-\frac{4v^2(2m+1)}{3\zeta_{0}(m+2)^2
[4+(m+2)\tilde{\Phi}]^2}-\frac{5(2m+1)}{4}\tilde{\Phi}
-\frac{4m+9}{8(m+2)}.\\\label{28b}
\end{eqnarray}
In case of viscous radiation like fluid, the location of source and
sink can be observed according to the sign of eigenvalues as well as
direction of the trajectories. We investigate stability of the
critical points corresponding to different values of $m$ and other
parameters. For $v=\sqrt{2/3}$ and $0.68$, sink $P_{d}^-$ lies in
the region with $q>0$ showing decelerated expansion of the universe
model (Figure \textbf{3}). The green region gradually increases by
increasing the values of $m$. We find accelerated expansion for more
realistic values of parameter $m$ approaching to unity dominated by
viscous radiation.
\begin{figure}\center
\epsfig{file=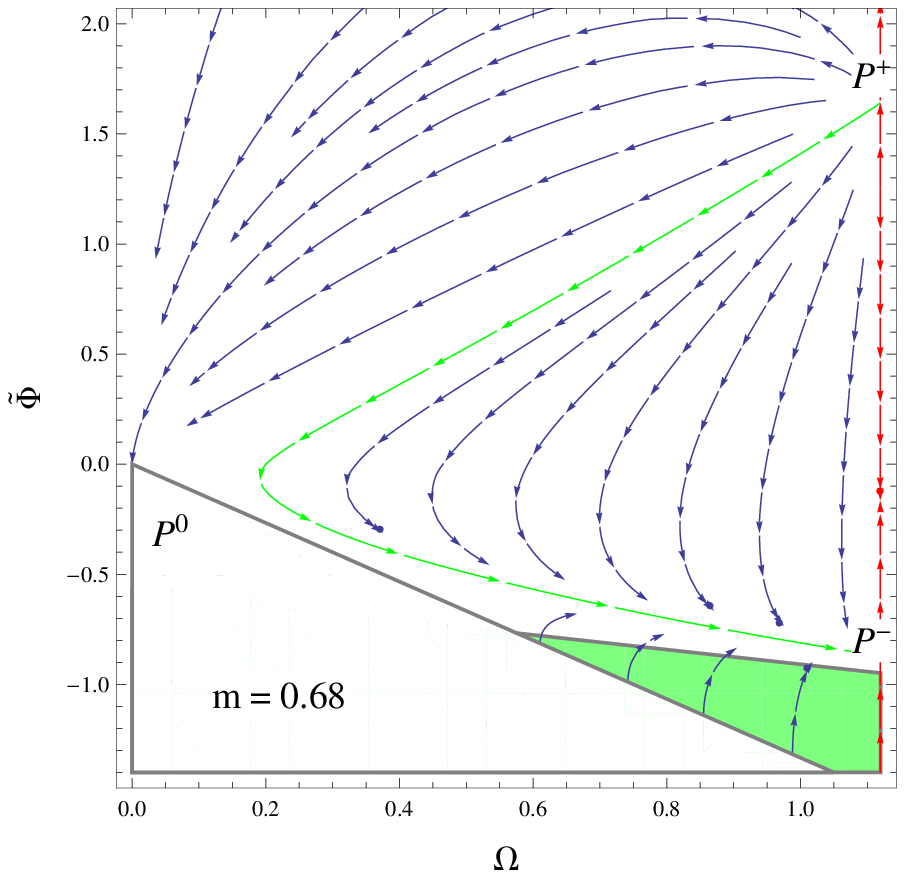,width=0.35\linewidth}\epsfig{file=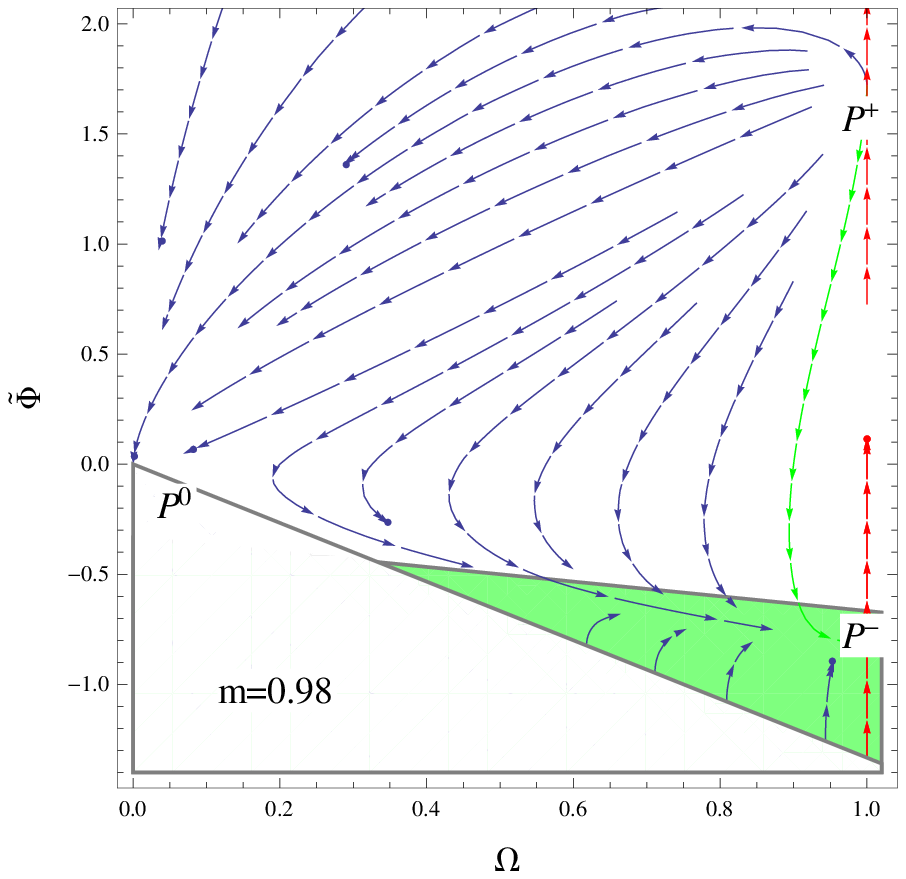,
width=0.35\linewidth}\epsfig{file=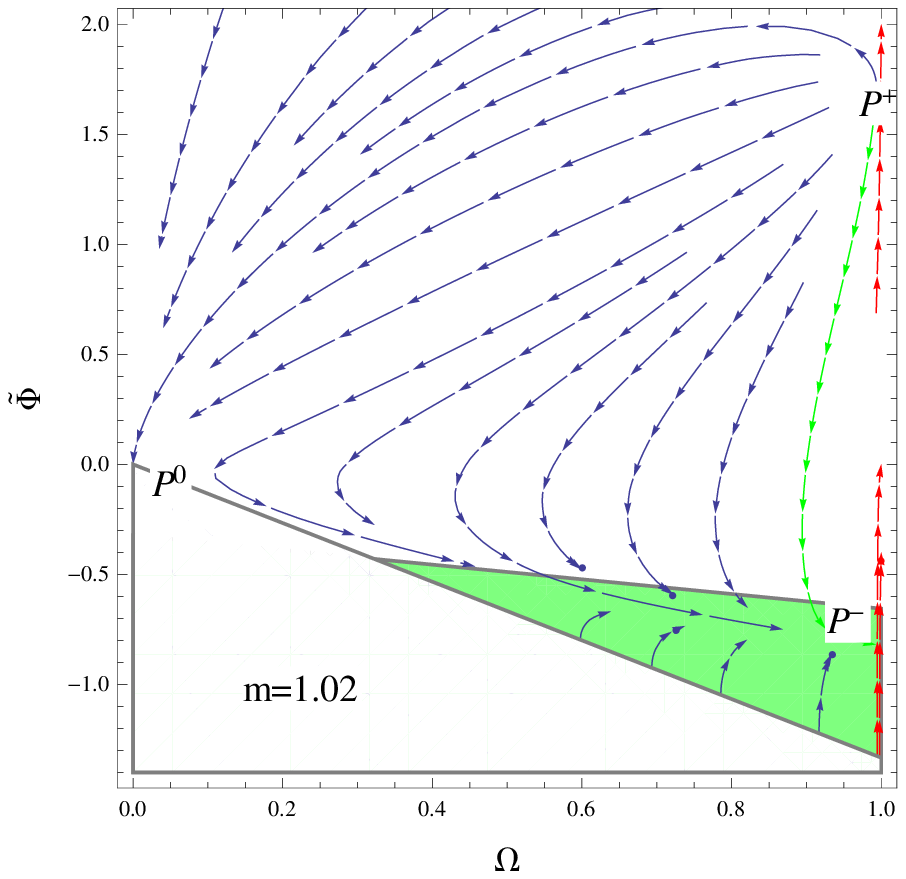,width=0.35\linewidth}
\caption{Plot for the phase plane evolution of viscous radiating
fluid with $w=4/3$, $v=\sqrt{2/3}$, $\zeta_{0}=1$ and different
values of $m$.}
\end{figure}

The behavior of critical point $P_{r}^*$ depends on the condition
(\ref{27aa}) and the choice of different parameters. If
Eq.(\ref{27aa}) holds, we observe that the points $P_{r}^*$ and
$P_{r}^-$ are stable attractors for $v=\bar{v}$,
$\zeta_{0}=\bar{\zeta_{0}}+1/10$ and different values of $m$. The
respective evolution plots are given in Figure \textbf{4}. It is
mentioned here that all choices of parameter $m$ show decelerated
expansion of the universe model with contributions coming from both
viscous radiation and non-viscous matter. We provide summary of our
results filled with viscous radiation like fluid in Table
\textbf{2}.
\begin{figure}\center
\epsfig{file=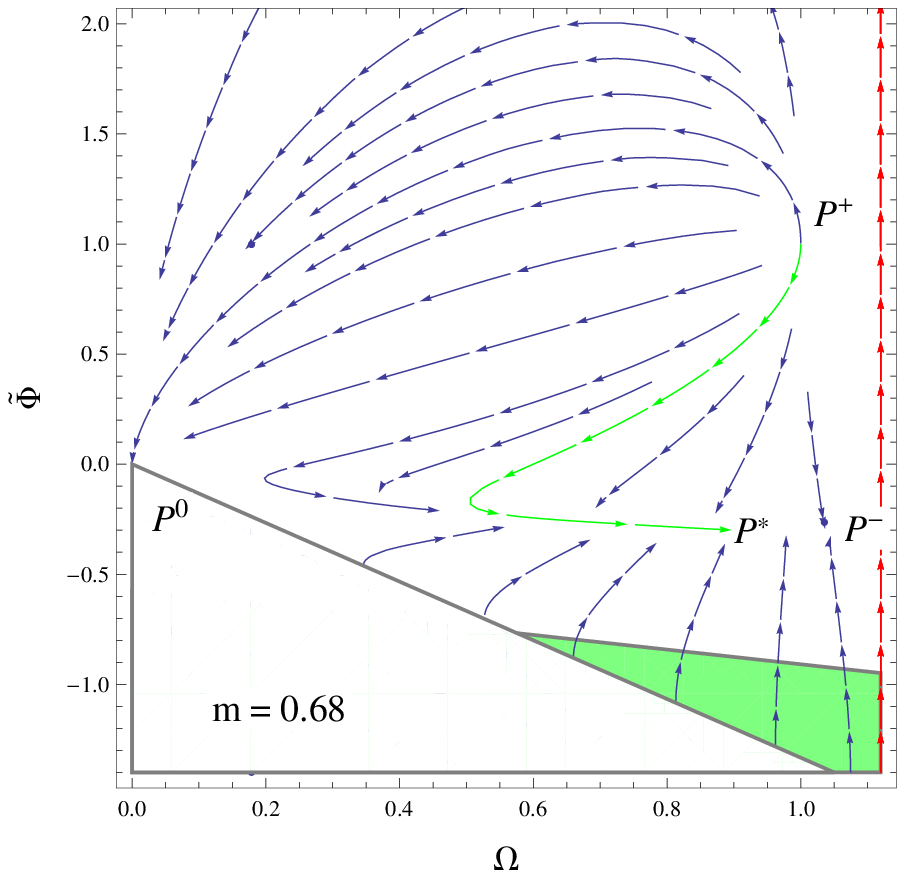,width=0.35\linewidth}\epsfig{file=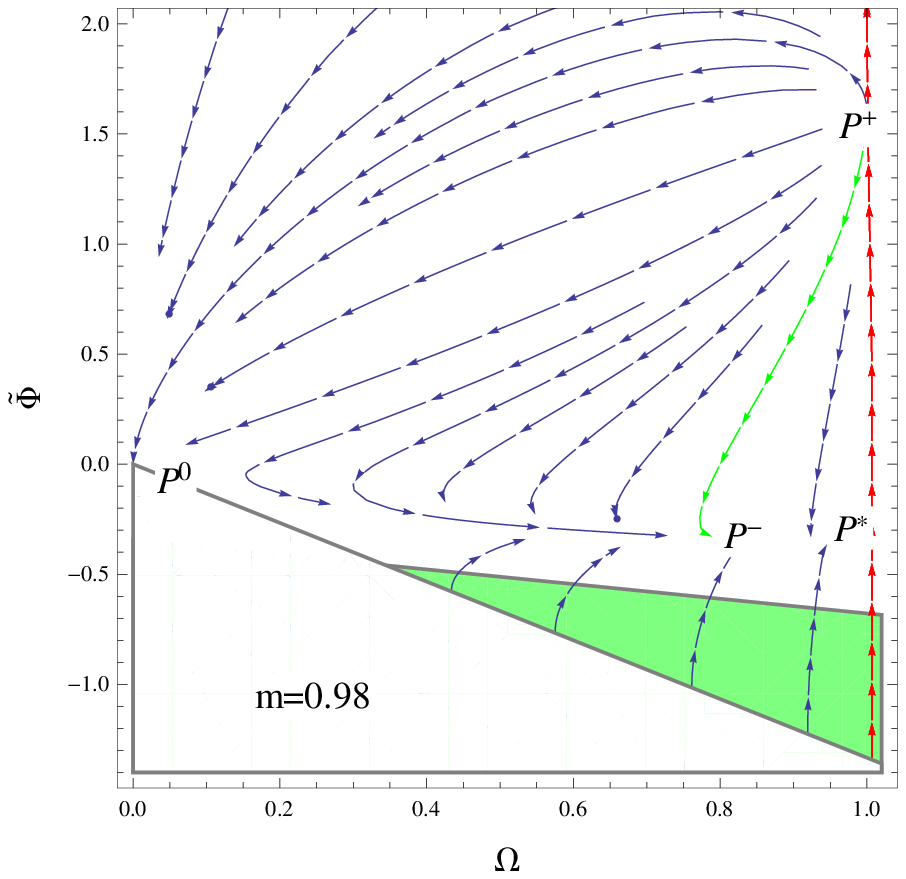,
width=0.35\linewidth}\epsfig{file=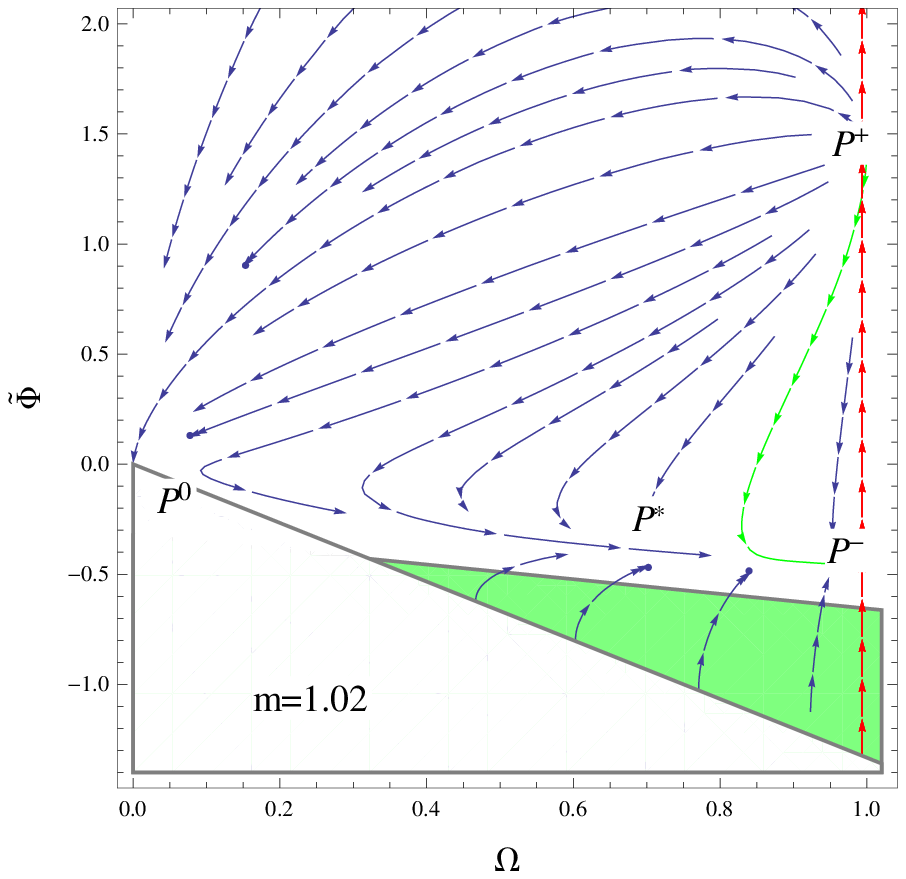,width=0.35\linewidth}
\caption{Plot for the phase plane evolution of viscous radiating
fluid with $w=4/3$, $v=\bar{v}$, $\zeta_{0}=\bar{\zeta_{0}}+1/10$
and different values of $m$.}
\end{figure}
\begin{table}[bht]
\textbf{Table 2:} \textbf{Stability Analysis of Critical Points for
Radiation Dominated Fluid}\\\\ \vspace{0.5cm} \centering
\begin{small}
\begin{tabular}{|c|c|c|c|}
\hline\textbf{Critical Point}&\textbf{If Eq.(\ref{27aa}) holds}&
\textbf{If Eq.(\ref{27aa}) does not hold}\\
\hline{\textbf{$P_{r}^0$}}&{Saddle}&{Saddle}\\
\hline{\textbf{$P_{r}^+$}}&{Source}&{Source}\\
\hline{\textbf{$P_{r}^-$}}&{Sink/Saddle}&{Sink}\\
\hline{\textbf{$P_{r}^*$}}&{Sink}&{-}\\
\hline
\end{tabular}
\end{small}
\end{table}

Here we also provide a comparison of our results with the work done
in literature. Chimento et al. \cite{b2} discussed the behavior of
homogeneous and isotropic universe model for both barotropic as well
as the ideal gas temperature. They investigated asymptotic stability
of the de Sitter and Friedmann solutions in which the former is
stable for bulk viscosity index having values less than unity and
the latter for the values greater than $1$. In our case, we consider
barotropic temperature. The bulk viscosity coefficient $\zeta_0=1$
as well as smaller values of parameter $m$ show that the stable
attractor lies in the region of decelerated expansion showing matter
dominated universe followed by viscous radiation. The region for
accelerated expansion tends to increase by increasing the values of
$m$ such that we investigate accelerated expansion (de Sitter
universe) for more realistic values of parameter $m$ approaching to
unity. For $\zeta_0>1$, we find stability of decelerated expansion
of the universe model with contributions coming from both viscous
radiation and non-viscous matter for all choices of parameter $m$.

\section{Power-Law Scale Factor}

In this section, we apply some assumptions on the scale factors
corresponding to the critical points. In this way, Eq.(\ref{13})
yields
\begin{equation}\label{28}
\dot{\Theta}=-\frac{1}{2}\left[\frac{3}{m+2}
+\frac{2m+1}{3}[(w-1)\Omega+\tilde{\Phi}]\right]\Theta^2.
\end{equation}
For $\Theta\neq0$, we obtain power-law scale factor whenever
$\frac{3}{m+2} +\frac{2m+1}{3}[(w-1)\Omega+\tilde{\Phi}]\neq0$.
Solving $\Theta=\frac{\dot{a}}{a}+\frac{2\dot{b}}{b}$ for $a(t)$ and
$b(t)$, we find the corresponding generic critical point as
\begin{equation}\label{29}
b^{(m+2)}=b_{0}^{(m+2)}(t-t_{0})^{\frac{2}{\frac{3}{m+2}+\frac{2m+1}{3}
[(w-1)\Omega_{c}+\tilde{\Phi_{c}}]}}.
\end{equation}
For exponentially expanding physical regions, the following
condition must be satisfied
\begin{equation}\label{30}
(1-w)\Omega_{c}- \frac{9}{(m+2)(2m+1)}>-\frac{wv^2}{k^2}\Omega.
\end{equation}
This condition does not hold in the physical phase space for
$v^2=k^2$. If $v^2>k^2$, the above inequality must be satisfied in
the following physical phase space region
\begin{equation}\label{31}
\frac{9}{(m+2)(2m+1)}\left[1-w\left(1-\frac{v^2}{k^2}\right)\right]^{-1}
<\Omega\leq\frac{3}{m+2}.
\end{equation}
It is mentioned here that the sign of the term $\frac{3}{m+2}
+\frac{2m+1}{3}[(w-1)\Omega+\tilde{\Phi}]$ is quite important to
evaluate different cosmological stages. If $\frac{3}{m+2}
+\frac{2m+1}{3}[(w-1)\Omega+\tilde{\Phi}]=0$, it corresponds to the
exponential expansion of the universe model. Also,
$\frac{3}{m+2}+\frac{2m+1}{3}[(w-1)\Omega+\tilde{\Phi}]\gtrless0$
yields accelerated expansion or contraction of the cosmological
model, respectively. If $v^2<k^2$, the possibility of having
accelerated expansion will narrow down and the green region will
disappear from the physical phase space. We plot the respective
results for power-law scale factor to explore different phases of
the universe model. Figures \textbf{5} and \textbf{6} show the
physical phase space region (above the white region) whereas green
and dark gray regions correspond to accelerated expansion and
contraction, respectively. For $v^2=k^2$, we find green region for
accelerated expansion which gets larger by increasing $m$. For
$v^2>k^2$, we find both expansion and contraction regions for the
cosmological model. In this case, the contraction decreases by
increasing $m$ while the green region becomes larger. Table
\textbf{3} shows the polynomial behavior of power-law scale factors
corresponding to different critical points with $\frac{3}{m+2}
+\frac{2m+1}{3}[(w-1)\Omega+\tilde{\Phi}]\neq0$.
\begin{table}[bht]
\textbf{Table 3:} \textbf{Power-law Scale Factors for Different
Critical Points}\\\\ \vspace{0.5cm} \centering
\begin{small}
\begin{tabular}{|c|c|c|c|}
\hline\textbf{Critical Point}&\textbf{Scale factors for $w=1$}&
\textbf{Scale factors for $w=4/3$}\\
\hline{\textbf{$P^0$}}&{$b_{0}^{(m+2)}(t-t_{0})^{\frac{2(m+2)}{3}}$}&{$b_{0}^{(m+2)}(t-t_{0})^{\frac{2(m+2)}{3}}$}\\
\hline{\textbf{$P^+$}}&{$b_{0}^{(m+2)}(t-t_{0})^{\frac{2}{\frac{3}{m+2}+\frac{2m+1}{3}
\tilde{\Phi}^{+}_{c}}}$}&{$b_{0}^{(m+2)}(t-t_{0})^{\frac{2}{\frac{2(m+5)}{3(m+2)}+\tilde{\Phi}^{+}_{c}}}$}\\
\hline{\textbf{$P^-$}}&{$b_{0}^{(m+2)}(t-t_{0})^{\frac{2}{\frac{3}{m+2}+\frac{2m+1}{3}
\tilde{\Phi}^{-}_{c}}}$}&{$b_{0}^{(m+2)}(t-t_{0})^{\frac{2}{\frac{2(m+5)}{3(m+2)}+\tilde{\Phi}^{-}_{c}}}$}\\
\hline{\textbf{$P^*$}}&{-}&{$b_{0}^{(m+2)}(t-t_{0})^{\frac{2(m+2)}{3}}$}\\
\hline
\end{tabular}
\end{small}
\end{table}
\begin{figure}\center
\epsfig{file=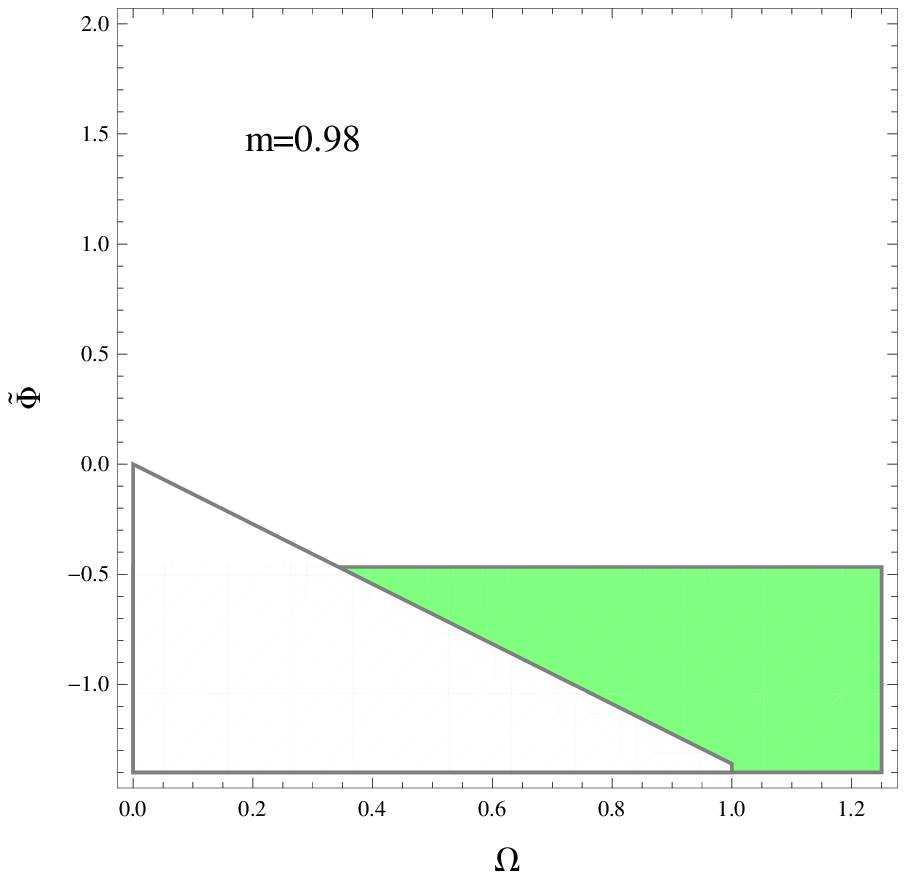,width=0.45\linewidth}\epsfig{file=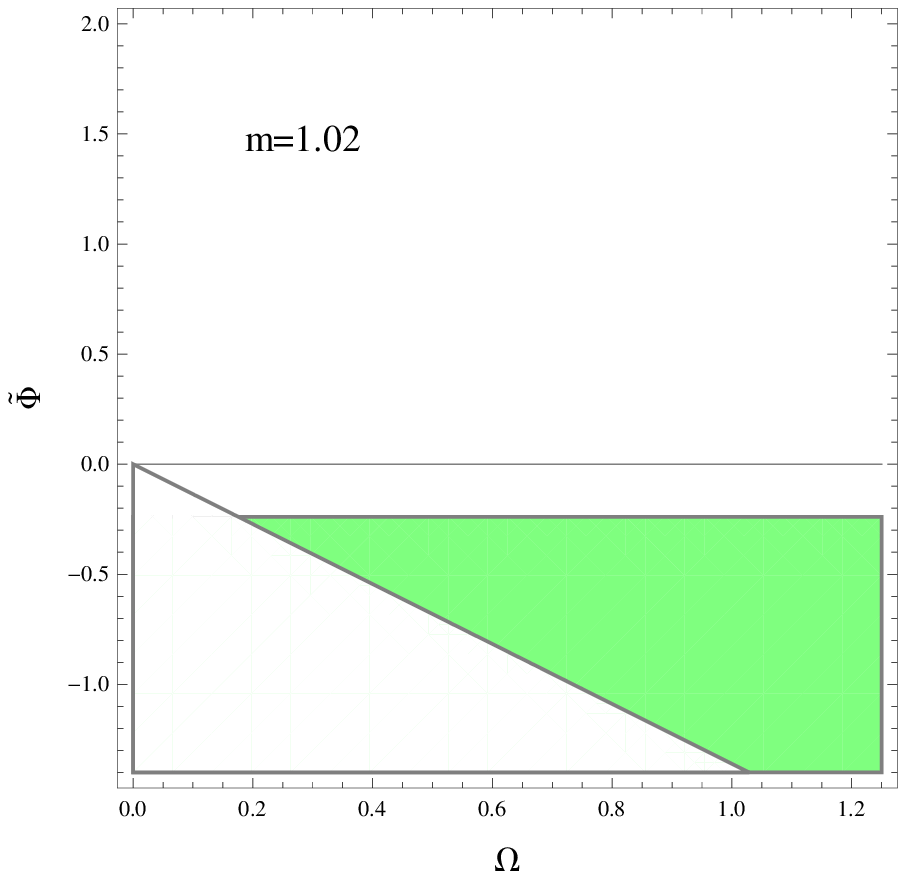,width=0.45\linewidth}
\caption{Plot of qualitative phase space analysis for power-law
scale factor with $v^2=k^2$.}
\end{figure}
\begin{figure}\center
\epsfig{file=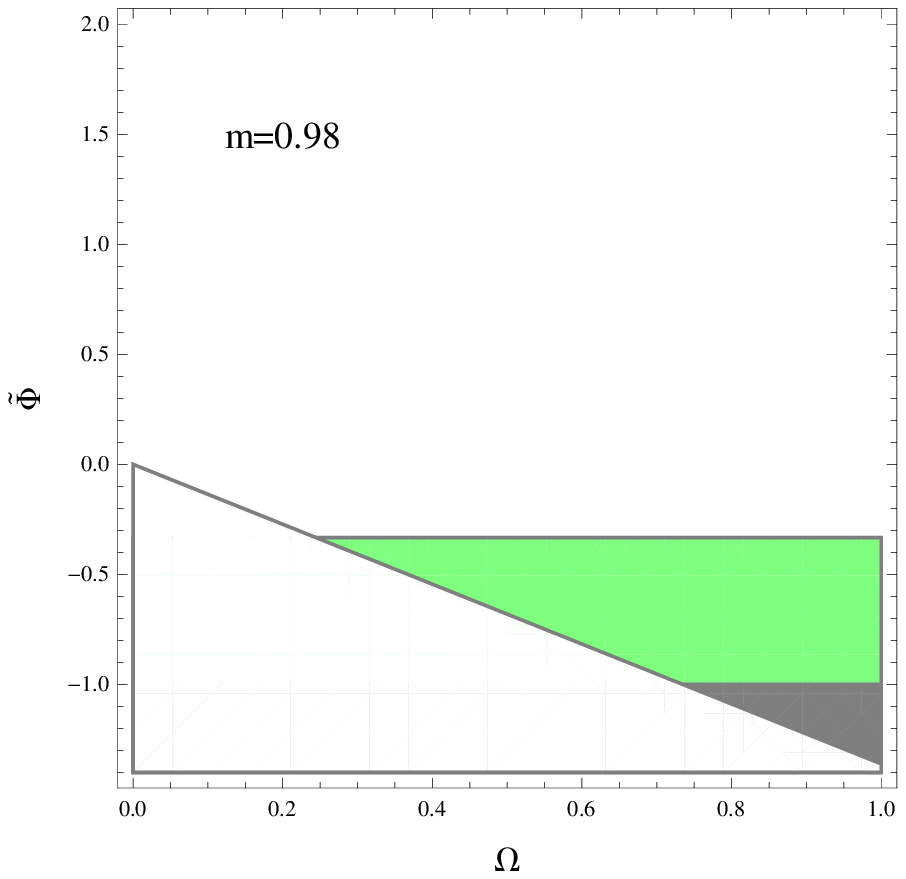,width=0.45\linewidth}\epsfig{file=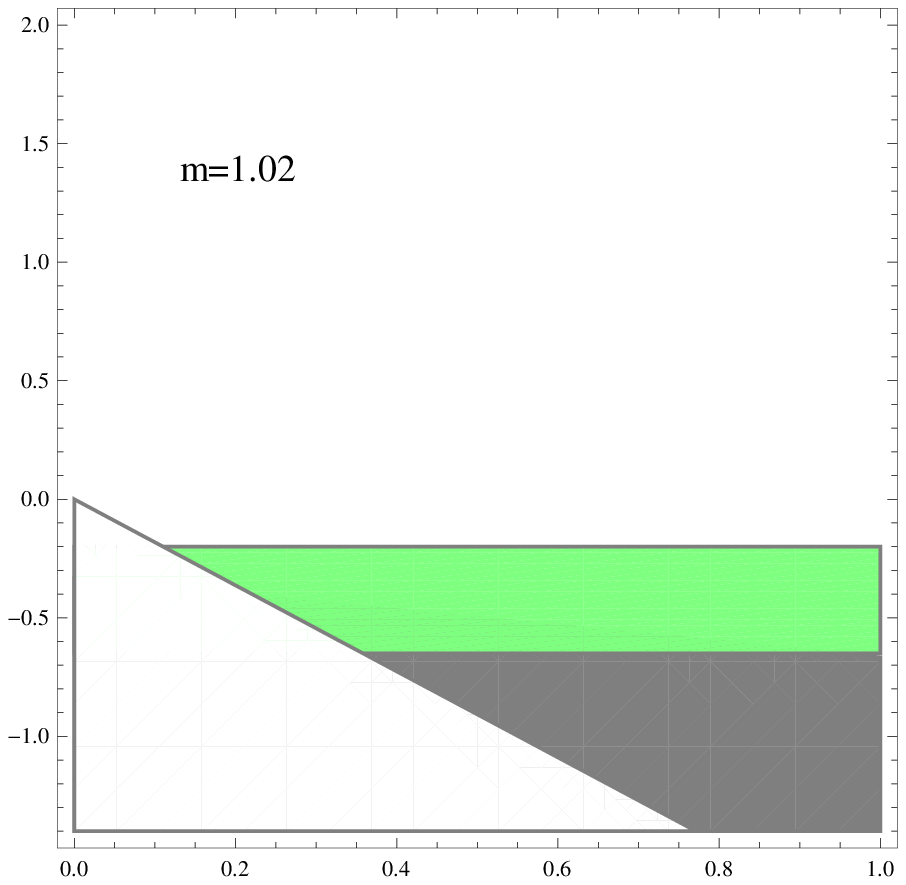,width=0.45\linewidth}
\caption{Plot of qualitative phase space analysis for power-law
scale factor with $v^2>k^2$. Green and dark gray regions indicate
the accelerated expansion and contraction of the universe model,
respectively.}
\end{figure}

\section{Summary}

This paper is devoted to study the phase space analysis for LRS BI
universe model by taking noninteracting mixture of dust like and
viscous radiation like fluids. This analysis has been proved to be a
remarkable technique for the study of stability of dynamical system.
An autonomous system of equations has been developed by defining
normalized dimensionless variables. In order to discuss stability of
the system, we have evaluated the corresponding critical points for
different values of the parameters. We have also calculated
eigenvalues which characterize these critical points. Moreover, we
have applied some assumptions on the scale factors to obtain
power-law scale factor whose behavior indicates the expansion or
contraction of the universe model. We summarize our results as
follows.

Firstly, we have discussed stability of critical points through
their eigenvalues corresponding to different values of $m$ for
pressureless fluid. It is found that the critical points $P^{+}_{d}$
and $P^{-}_{d}$ correspond to source (unstable) and sink (stable),
respectively (Figures \textbf{1} and \textbf{2}). The white region
shows the universe models with a negative entropy production rate
which diverges on its boundary. It is mentioned here that
trajectories in its neighborhood are not attracted towards the
boundary showing its significant role to keep the models away from
divergence. The green region corresponds to the accelerated
expansion of the universe. It is found that the point $P_{d}^-$ is a
global attractor in the physical phase space region which leads to
an expanding model dominated by viscous matter for $m$ approaching
to unity and $v^2=k^2=1$ while $m=0.68$ corresponds to deceleration
of the respective model. For $v^2=k^2=0.04$, all choices of $m$ show
decelerated expansion dominated by matter.

Secondly, we have studied stability of the critical points in a
viscous radiation like fluid. In this case, the critical points
$P_{r}^+$ and $P_{r}^-$ correspond to source and sink, respectively.
If Eq.(\ref{27aa}) holds, we have found that the behavior of
$P_{r}^-$ is not fixed rather depends on the values of different
parameters. In the case of viscous radiation, we have emphasized on
the fact that any trajectory starting from a neighborhood of
$P^+_{r}$ will go through the following stages in physical phase
space region: (i) source $P_{r}^+$ corresponds to a radiation
dominated era, (ii) saddle $P_{r}^0$ showing a matter dominated era,
(iii) decelerated expansion (sink $P_{r}^-$ or $P_{r}^*$). It is
found that stable solutions exist for noninteracting fluids in the
presence of nonlinear bulk viscosity for $m$ closer to unity which
show accelerated expansion of the universe model. If Eq.(\ref{27aa})
does not hold, the universe model is in decelerating era for all the
choices of $m$. It is worth mentioning here that $m=0.98,~1.02$ are
more acceptable values for phase space analysis of LRS BI universe
model.

We have also obtained power-law scale factor whose behavior
indicates expansion or contraction of the universe model for
different values of $m$ and the other parameters. Figures \textbf{5}
and \textbf{6} show the physical phase space region (above the white
region) whereas green and dark gray regions correspond to
accelerated expansion and contraction, respectively. The boundary
between green and dark gray regions represents exponential expansion
of the universe model. For $v^2=k^2$, it is found that the green
region for accelerated expansion gets larger by increasing $m$. For
$v^2>k^2$, we find both expansion and contraction regions for the
respective cosmological model. In this case, the contraction
decreases by increasing $m$ while the green region becomes larger.
We conclude that our analysis does not provide a complete immune
from fine-tuning because the exponentially expanding solution occurs
only for a particular range of parameters.

\section*{Acknowledgement}

We would like to thank the Higher Education Commission, Islamabad,
Pakistan for its financial support through the \emph{Indigenous
Ph.D. Fellowship, Phase-II, Batch-III}.

\end{document}